\documentclass[journal]{IEEEtran}

\usepackage{amsmath,amssymb,epsfig}
\usepackage{cite,footnote,authblk}
\usepackage{color,hyperref}
\hypersetup{
    colorlinks,%
    citecolor=blue,%
    filecolor=blue,%
    linkcolor=blue,%
    urlcolor=blue
}
\definecolor{bluecolor}{rgb}{0,0.,1.}

\definecolor{redcolor}{rgb}{.7,0.,0.}

\newcommand{\pr}[1]{\left( #1\right)}
\newcommand{\prr}[1]{\left[ #1 \right]}
\newcommand{\es}[1]{\begin{equation}\begin{split}#1\end{split}\end{equation}}

\newcommand{\R}{\mathbb{R}}

\newcommand{\V}{\mathcal{V}}
\newcommand{\M}{\mathcal{M}}

\newcommand{\rr}{\mathbf{r}}
\newcommand{\dd}{\textrm{d}}
\newcommand{\vv}{\hat{\mathbf{v}}}

\usepackage{fancyhdr}
\begin{document}

\title{Connectivity of confined 3D Networks with Anisotropically Radiating Nodes}
\author[1,2]{Orestis Georgiou\thanks{orestis.georgiou@toshiba-trel.com}}
\author[2]{Carl P. Dettmann}
\author[3]{Justin P. Coon}
\affil[1]{Toshiba Telecommunications Research Laboratory, 32 Queens Square, Bristol, BS1 4ND, UK}
\affil[2]{School of Mathematics, University of Bristol, University Walk, Bristol, BS8 1TW, UK}
\affil[3]{Department of Engineering Science, University of Oxford, Parks Road, OX1 3PJ, Bristol, UK}
\maketitle

\IEEEpeerreviewmaketitle

\pagestyle{plain}

\thispagestyle{fancy}

\begin{abstract}
Nodes in ad hoc networks with randomly oriented directional antenna patterns typically have fewer short links and more long links which can bridge together otherwise isolated subnetworks. 
This network feature is known to improve overall connectivity in 2D random networks operating at low channel path loss. 
To this end, we advance recently established results to obtain analytic expressions for the mean degree of 3D networks for simple but practical anisotropic gain profiles, including those of patch, dipole and end-fire array antennas.
Our analysis reveals that for homogeneous systems (i.e. neglecting boundary effects) directional radiation patterns are superior to the isotropic case only when the path loss exponent is less than the spatial dimension.
Moreover, we establish that ad hoc networks utilizing directional transmit and isotropic receive antennas (or vice versa) are always sub-optimally connected regardless of the environment path loss.
We extend our analysis to investigate boundary effects in inhomogeneous systems, and study the geometrical reasons why directional radiating nodes are at a disadvantage to isotropic ones.
Finally, we discuss multi-directional gain patterns consisting of many equally spaced lobes which could be used to mitigate boundary effects and improve overall network connectivity.
\end{abstract}


\section{Introduction \label{sec:intro}}

Wireless ad hoc networks do not rely on any pre-existing infrastructure such as routers or access points and so can be deployed on the fly \cite{cordeiro2011ad}.
Equipped with multihop relaying and signal processing capabilities, they can self-organize and dynamically optimize network performance, traits which are becoming increasingly useful in sensor and vehicular network applications \cite{hartenstein2008tutorial}, including, \textit{inter alia}, exploration and environmental monitoring over extended 3D regions \cite{pompili2006deployment,savazzi2013ultra}, 
disaster detection and/or search-and-rescue operations in hazardous/disaster relief areas \cite{reich2006robot,aziz2011managing}, swarm robotics \cite{mohan2009extensive},
road safety message dissemination, traffic management and dynamic route
planning \cite{Kafsi08vanetconnectivity,viriyasitavat2011dynamics}.
Commonality in these applications can be found in that the number and distribution of nodes in the networks is often random, as was realized and studied by Gupta and Kumar in 1998 \cite{gupta1998critical}.
From a communications perspective, understanding the connectivity properties of random networks\footnote{A plethora of relevant problems and solutions can be found in the mathematical literature under random graph theory \cite{penrose2003random}, and in physics under percolation theory \cite{bollobas2006percolation,albert2002statistical}.} has ever since been of paramount importance as it can lead to improved network design, protocols and deployment methodologies \cite{romer2004design,ravelomanana2004extremal}.

It is often assumed, that when deployed, ad hoc networks will be well connected. 
To date, many works have challenged this assumption and have theoretically investigated a number of network features and variants.
Most however adopt one or more of the following assumptions: the network resides in an extended two dimensional domain\footnote{Some works consider bounded domains but scale volume exponentially with the number of nodes, thus ignoring any boundary effects \cite{mao2012towards}.}, nodes are isotropically radiating, links between nodes are formed deterministically according to a fixed range (i.e. unit-disk type connections), and/or the number of nodes $N \to \infty$.
In what follows we will lift all of these assumptions.

We are specifically interested in the effect of \textit{randomized beamforming} strategies\footnote{In randomized beamforming, each anisotropically radiating node selects a boresight direction randomly and independently on the unit sphere.} which are known to improve network connectivity at low path loss exponents. 
How this improvement is achieved was first addressed in \cite{bettstetter2005does}, and later in \cite{koskinen2006analytical} where it was argued that randomized beamforming cannot be said to strictly improve/degrade connectivity.
To this end, it was numerically estimated in \cite{zhou2009connectivity} (and similar papers by the same authors) that the critical path loss exponent below which improvements are observed is $3$.
This was analytically pushed down to $2$ in \cite{coonanisotrop} where it was also shown that this number is independent of the small-scale fading model used.
Finally, the possibility of using multi-directional antennas was proposed and studied in \cite{xu2012connectivity}, and although unmotivated, was reported to enhance connectivity at low path loss.

It is not surprising however that most (if not all) relevant studies are restricted to two dimensional networks.
This partial understanding is what motivates the current investigation where we consider finite and confined three dimensional networks with anisotropically radiating nodes, that connect in probability space using well established statistical fading models.
To this end, we provide general analytic formulas for the connectivity mass\footnote{The connectivity mass (defined below) is a measure of the likelihood that any given node will be connected, and is related to several other network connectivity observables.} of several simple but practical radiation pattern approximations (including those of patch, dipole and end-fire array antennas) and conclusively show \textit{when and how} randomized beamforming of anisotropically radiating nodes can improve or worsen the connectivity of ad hoc networks\footnote{In all cases, the gain is properly normalized as to ensure a fair comparison.}.
Namely, we find that in the absence of boundary effects, directional antennas yield superior performance when the path loss exponent $\eta$ is less than the spatial dimension $d$, and inferior when $\eta>d$. 
Moreover, when $\eta=d$, network connectivity is found to be invariant to the antenna gain details.
This simple and attractive picture is however radically different in confined spaces. 
We show that in the presence of boundaries, the advantages of directional antennas are significantly undermined due to the existence of `blind spots', which effectively decrease the network mean degree and increase the likelihood of node isolation.
Therefore, we propose and investigate multi-directional radiation patterns as a means to mitigate boundary effects.

The paper is structured as follows: 
Sec.~\ref{sec:setup} introduces the system set-up and all relevant parameters and assumptions. 
Sec.~\ref{sec:NC} discusses various network observables and identifies the connectivity mass $\mathsf{M}$ as a key quantity of interest. 
Sec.~\ref{sec:funct} investigates the connectivity mass for homogeneous systems (i.e. ignoring boundary effects) and derives analytic expressions for $\mathsf{M}$ for simple but practical gain profiles which are then verified through computer simulations.
Based on these expressions, Sec.~\ref{sec:DS} reveals that the connectivity properties of ad hoc networks scale with the solid angle over which the gain is concentrated on.
Sec.~\ref{sec:BE} examines the effect of boundaries in inhomogeneous systems and identifies the weaknesses of directional patterns.
Sec.~\ref{sec:multi} proposes a multi-directional solution to mitigate boundary effects and investigates the optimal radiation pattern for a rectangular cuboid domain. 
Finally, Sec.~\ref{sec:conc} summarises and discusses the main results and highlights some ideas and challenges for future research.

\section{Network and System Model \label{sec:setup}}

We begin our discussion by describing the system set-up and all relevant parameters and assumptions.
We consider a network consisting of $N$ identical nodes with locations $\rr_i$ for $i=1,\ldots,N$, chosen randomly inside a three dimensional domain $\V\subseteq\R^3$ of volume $V$.
The density of nodes is assumed uniform and is given by $\rho=N/V$. 
Such spatial node distributions are often used to model vehicular, ad hoc, and wireless mesh sensor networks, which are either dynamically evolving or are free from any pre-existing infrastructure \cite{cordeiro2011ad,hartenstein2008tutorial}.
Assuming negligible inter-node interference, we say that two nodes $i$ and $j$ separated by a distance $r_{ij}=|\rr_i - \rr_j|\geq 0$ are connected if the channel between them supports a rate of at least $\wp$
\es{
H_{ij}=P(\log _2 (1+ \textrm{SNR} \cdot |h|^2) > \wp )
,
\label{H1}}
where SNR denotes the long-term average received signal-to-noise ratio and $h$ is the channel transfer coefficient for single input single output (SISO) antenna systems. 
For the sake of simplicity, we will assume only small scale scattering effects and thus adopt a Rayleigh fading model where $|h|^2$ is modelled as an exponentially distributed random variable \cite{tse2005fundamentals}. 
It should be noted however that more exotic fading models such as the two-wave with diffuse power (TWDP) \cite{durgin2002new} which can approximate channels with arbitrary combinations of specular and diffuse components, offer little additional insight to the present discussion as our approach will be concentrated on the antenna radiation patterns rather than the detailed fading parameters \cite{coonanisotrop}.

Assuming lossless antennas with equal transmit and receive performances we have from the Friis transmission formula that
\es{
\textrm{SNR}\propto  G_i G_j  r_{ij}^{-\eta}
\label{snr}}
where $\eta$ is the path loss exponent\footnote{Typically $\eta=2$ corresponds to propagation in free space but for cluttered environments it is observed to be $\eta>2$.
Values of $\eta<2$ have been reported, typically for indoor environments, for example in grocery stores \cite{seidel1992914}.},
$G_{i}$ is the gain of the antenna at node $i$ observed in the direction of node $j$ and
$G_{j}$ is the gain of the antenna at node $j$ observed in the direction of node $i$.
Hence, we may express the pair connectedness probability of nodes $i$ and $j$ as
\es{
H_{ij}=\exp\pr{- \frac{\beta r_{ij}^{\eta}}{G_i G_j}},
\label{H}}
where $\beta$ defines the characteristic connection length 
\es{r_0= \pr{\frac{\beta}{G_i G_j}}^{-1/\eta}
,}
and depends on for example the transmission wavelength, signal power, \textit{etcetera}.

\begin{figure}[t]
\centering
\includegraphics[scale=0.25]{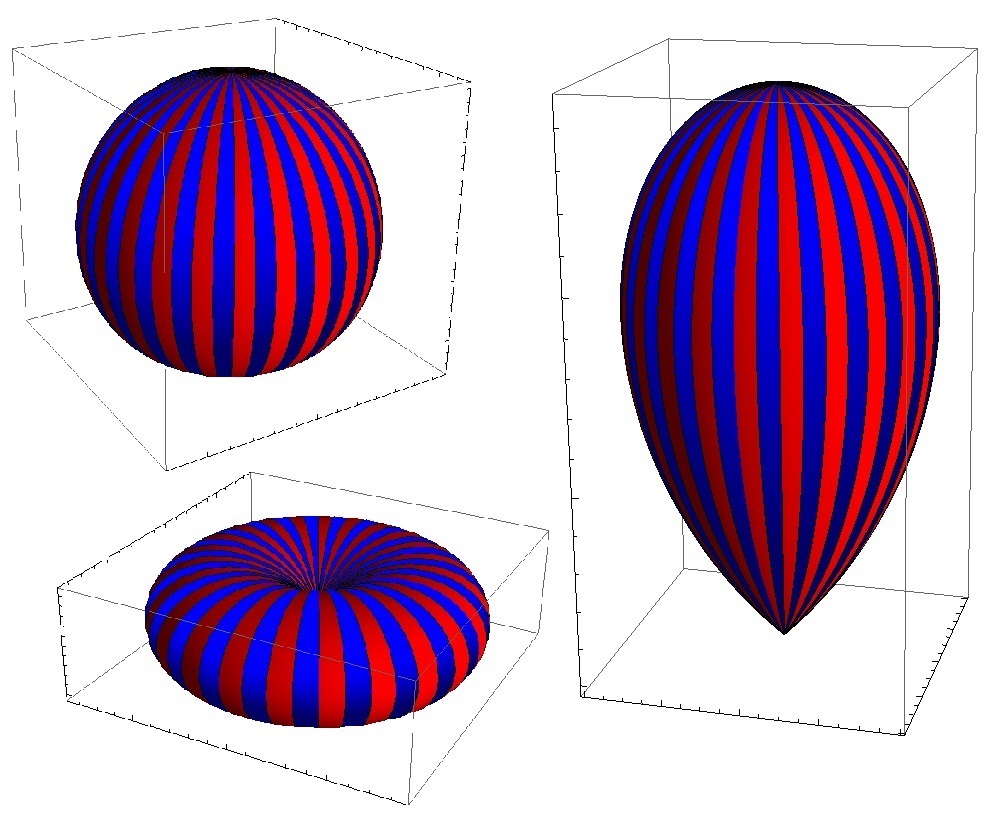}
\caption{Simplified gain patterns of a patch, dipole, and horn or end-fire array antennas.}
\label{fig:pat}
\end{figure}

The gain functions represent the ratio between the signal intensity in a given direction, and the signal intensity had the same power been radiated isotropically.
In order to keep the mathematics tractable, we will ignore small sidelobes and backlobes (as done in many other studies) and mostly restrict our analysis to rotationally symmetric gain patterns (i.e. surfaces of revolution) about some orientation unit vector $\vv$.
It follows that isotropic radiation patterns have a constant gain $G=1$, while anisotropic ones are functions of the polar angle $\theta$ about $\vv$, appropriately normalized by the condition
\es{
\int_{0}^{2\pi }\!\! \int_{0}^{\pi} G(\theta) \sin \theta \dd \theta \dd \phi
=4\pi
.\label{norm}}
At a later stage, we will also consider \textit{multi-directional} radiation patterns of the form
\es{
G(\theta,\phi)= \sum_{k=1}^n g_k(\theta^{(k)}) 
\label{multi},}
where the $g_k$ characterize the gain function for the $k$-th directional lobe, and $\theta^{(k)}$ is the angle measured from a collection of unit vectors $\varTheta = \{\vv^{(k)}, \textrm{ for } k=1,\ldots,n\}$.
Note that multi-directional radiation patterns are typically not surfaces of revolution, but their individual components $g_k$ are.
We adjourn further development of this notation until the relevant subsections.

Complicated radiation patterns can be achieved using beamforming techniques \cite{ramanathan2001performance}.
In the simplest case this would involve a number of isotropic radiating antenna elements arranged in a linear, circular or planar array with variable transmit and receive powers and appropriately tuned phase shifts.
If done correctly, the constructive/destructive electromagnetic interference results in the desired radiation pattern.
Achieving perfect beamforming across large networks however is an idealized and relatively costly system requirement.
Therefore, we adopt here a simpler strategy by choosing the antenna orientation vector $\vv$ of each node at random
thus offering a simple and practical deployment method at low hardware complexity and minimal communication overhead.

In summary, our system model, as defined above, has three sources of randomness: random node positions, random antenna orientations, and random pair connection probabilities according to the channel fading model. 
In what follows, we investigate the network connectivity properties for simple but practical radiation patterns characteristic of relatively cheap and readily available antennas, e.g., patch, dipole, and horn or end-fire arrays (see Fig.~\ref{fig:pat}).


\section{Network Connectivity and Connectivity Mass \label{sec:NC}}

There exist a plethora of measures of the connectivity properties of a complex network \cite{albert2002statistical}. 
These include for example clustering statistics, network modularity measures, the number of independent paths, algebraic connectivity, etc.
Each of these measures offers different information and one must choose wisely which ones are useful to the intended application.
Here, we restrict our attention to three closely related observables, namely we study 1) the  pair formation probability $p_2$, 2) the degree distribution $d(k)$, and 3) the probability that a random network with randomly oriented antennas is fully connected $P_{fc}$. 
In what follows, we will argue that all three observables can be effectively characterized through the connectivity mass $\mathsf{M}$ thus rendering it the key metric of interest which we will then study in some detail in the subsequent sections.

\subsection{Pair Formation Probability}

The probability that node $i$ situated at $\rr_i$ connects with a randomly chosen node $j$ is obtained by averaging over all possible node positions $\rr_j\in\V$ and all possible antenna orientations
\es{
H_i =\frac{1}{4\pi V} \int \! \int_\V H_{ij}  \dd \rr_j \dd \Omega_j 
,\label{Hi}}
where $\dd \Omega = \sin\vartheta\dd\vartheta \dd\varphi$ denotes the differential solid angle in spherical coordinates.
Note that we will use curly symbols $(\vartheta,\varphi)$ for orientation coordinates, and normal ones $(\theta,\phi)$ for position coordinates.
Integrating \eqref{Hi} once more gives the probability that two randomly selected nodes connect to form a pair
\es{
p_2 &=  \frac{1}{(4\pi V)^2} \!\! \int \!\! \int_{\V^{2}} H_{ij}  \dd \rr_i \dd \rr_j \dd \Omega_i \dd \Omega_j \\
&= \! \frac{1}{4\pi V} \!\! \int \!\! \int_\V H_{i}  \dd \rr_i \dd \Omega_i 
.}

\subsection{Degree Distribution}

Since node locations and orientations are independent, the probability that node $i$ has degree $k$ (i.e. connects with exactly $k$ other nodes) is given by the binomial distribution
\es{
d_i(k)= \binom{N-1}{k} H_i^{k}(1-H_i)^{N-1-k}
.
\label{bin}}
If $N$ is large and $H_{i}$ is small, \eqref{bin} can be well approximated by the Poisson distribution
\es{
d_i(k)\approx \frac{\mu_{i}^{k}}{k!} e^{-\mu_i}, \qquad D_{i}(k)=\sum_{m=0}^{k}d_{i}(m)
\label{nd},}
where $\mu_i= (N-1)H_i$, and $D_{i}(k)$
is the corresponding cumulative distribution function.
The Poisson approximation is justified here if $V\gg 1$, thus making $H_{i}$ small\footnote{This is due to division by $V$ in \eqref{Hi}.}.
Finally, to obtain the degree distribution we average over all possible node positions $\rr_i\in\V$ and all possible antenna orientations to obtain
\es{
d(k&)= \frac{1}{4\pi V} \int \!\!  \int_\V d_i(k) \dd \rr_i \dd \Omega_i \\
&= \frac{1}{4\pi V} \int \!\!  \int_\V \frac{\mu_{i}^{k}}{k!} e^{-\mu_i} \dd \rr_i \dd \Omega_i
.\label{dk}}
The average number of nodes connected to a typical node in the network is called the mean degree and is simply given by
\es{
\mu = \frac{1}{4\pi V} \int \!\! \int_\V \mu_{i}  \dd \rr_i \dd \Omega_i = (N-1)p_2
.}

\subsection{Full Connectivity}

A network is said to be fully connected if any node can communicate with any other node in a multihop fashion.
While a very strong measure of connectivity, $P_{fc}$ is compatible with delay and/or disruption tolerant networking recently popularized by the Defence Advanced Research Projects Agency (DARPA) in an attempt to increase wireless network reliability and prevent disruption due to: radio range, node sparsity, energy resources, attack, noise, etc. \cite{farrell2006tcp,fall2008dtn}.
It should be noted that the commonly discussed metric $P(\text{path})$ defined as the percentage of nodes that are connected via a multi-hop fashion is equivalent to $P_{fc}$ at high node densities.

For isotropic radiation ($G=1$), a theory for $P_{fc}$ was recently developed in \cite{coon2012full} for arbitrary dimension $d\geq1$ using an exact cluster expansion approach derived from statistical physics. 
We briefly highlight the key findings of \cite{coon2012full} as they are implicitly relevant in understanding the connectivity aspects of networks with anisotropically radiating nodes.
Their main result was a general formula expressing $P_{fc}$ at high node densities as the complement of the probability of an isolated node
\es{
P_{fc}&=1-\rho \int_{\V}e^{-\rho \int_\V H_{ij} \dd \rr_j} \dd \rr_i \\
&\approx 1- \rho \sum_{B} \textbf{G}_B \textbf{V}_B e^{-\rho \textbf{M}_B}
,\label{Pfciso}}
where the spatial $\dd \rr_i$ integral is approximated in the second line by a sum of separable contributions. 
Namely, for $d=3$, the domain $\V$ was partitioned (see also \cite{khalid2013connectivity} for a simple two dimensional square, and \cite{coon2012connectivity} for a three dimensional `house' domain) into a \textit{bulk} component, a \textit{surface area} component, and a number of \textit{edge} and \textit{corner} components.
Each component was then integrated on its own giving different volume $\textbf{V}_B$ and geometrical $\textbf{G}_B$ factors.
Most importantly however, to each component corresponds a different \textit{connectivity mass}
\es{
\textbf{M}_B = \omega_{B} \int_0^\infty r^{d-1} H(r) \dd r
, \label{Miso}} 
where $\omega_B$ is the solid angle subtended by boundary component $B$. For example $\omega_B= 4\pi, 2\pi, \pi$, and $\pi/2$ for the bulk, the surface area, a right angled edge, and right corner respectively.
The novelty of \eqref{Pfciso} lies in the logical decomposition of the domain into components of different full connectivity importance at different densities.
That is, at low to medium densities $\rho$, the bulk component dominates $P_{fc}$, followed by the surface area component at medium values of $\rho$, the edges at higher $\rho$, and finally corners at very high densities (see Fig. 5 of \cite{coon2012full}). 
Significantly, the asymptotic behaviour of $P_{fc}$ as $\rho \to \infty$ is characterized by the sharpest corner of the domain where $\omega_B$ is smallest.

Equation \eqref{Pfciso} can be generalized for anisotropic radiation patterns and can account for the randomly oriented nodes in a straightforward way as follows
\es{
P_{fc}&=1- \frac{\rho}{4\pi} \int \!\! \int_{\V}    e^{-\frac{\rho}{4\pi} \int \!\!\int_\V H_{ij}\dd \rr_j \dd \Omega_j } \dd \rr_i \dd \Omega_i 
\label{Pfc}
.}
There are several important observations to be highlighted here.
Firstly, we notice that at high node densities $P_{fc} = 1- N d(0)$ and therefore the two observables are intimately related and can be studied simultaneously.
Secondly, if $\V=\R^3$, the system becomes homogeneous and isotropic\footnote{Isotropic here refers to the whole network (i.e. appears uniform in all orientations) and not the radiation pattern of each individual node.} and therefore the $i$th node's position and orientation is arbitrary such that \eqref{Pfc} simplifies to
\es{
P_{fc}=1- N  e^{- \rho \mathsf{M} } 
\label{isoP},}
as there are no boundary components and the connectivity mass is given by
\es{
\mathsf{M}=\frac{1}{4\pi} \int \!\!\int_{\R^3} \exp\pr{-\frac{\beta r_j^\eta}{G_i G_j}}\dd \rr_j \dd \Omega_j
.\label{isoM}} 
Significantly, in this case we have that the pair formation probability is simply $p_2 = H_i = \mathsf{M}/V$, and the mean degree is $\mu= \rho \, \mathsf{M}$ when $N\gg1$.
Note that \eqref{isoP} and \eqref{isoM} also hold for any translationally or rotationally invariant surface in $\R^3$. 
One such example is if the nodes are situated on the surface of a sphere.
In conclusion, for a homogeneous system, $\mathsf{M}$ is the key observable of interest as it contains information about $P_{fc}$, $p_2$, $\mu$, and $d(0)$. Furthermore, $\mathsf{M}$ acts as a proxy to $d(k)$ which is then related to the $k$-connectivity \cite{georgiou2013k}; a strong measure of reliability and robustness.

Alternatively, if the domain is finite $\V\subset\R^{3}$ (and not translationally or rotationally invariant), the system becomes inhomogeneous and non-isotropic, and hence boundary effects can have a significant impact \cite{coon2012impact}.
In analogy to \eqref{Pfciso} we therefore expect $P_{fc}$ in the high density asymptotic limit to be dominated by contributions where the integrals in the exponents of \eqref{Pfc} are small, i.e. when $\rr_i$ is situated near the sharpest corner of $\V$ \textit{and} is also oriented such that its main connectivity beam is steered outside of $\V$.

Finally, we remark that contrary to $P_{fc}$ and $d(k)$, the pair formation probability $p_2$ and the mean degree $\mu$ are \textit{local} observables involving only $2$ nodes (rather than \textit{global} ones involving $N\gg1$ nodes\footnote{Intuitively, this is why the position and orientation integrals of node $j$ appear in the exponents of \eqref{dk} and \eqref{Pfc}.}).
Consequently, if the typical system size is much larger than the typical connectivity range $r_0$, we expect $p_2$ and $\mu$ to be very much insensitive to the domain shape details leading to $p_2 \lesssim \mathsf{M}/V$ and $\mu \lesssim \rho \, \mathsf{M}$.
This is because $H_i$ is approximately constant for the majority of node positions (away from the domain boundary) and orientations, and decreases (approximately linearly) when closer than $\sim r_0$ to the boundary.
We will confirm this expectation later through computer simulations (see Fig.~\ref{fig:num}).

In this section we have argued that the connectivity mass is a key metric which characterizes network connectivity in the high density limit.
Our aim in what follows is to analyse and understand the connectivity mass for different anisotropic radiation patterns and thus offer intelligent and useful design recommendations which improve connectivity of communication networks. 
We begin with the simple case of a homogeneous system i.e. ignoring any boundary effects.

\begin{figure}[t]
\centering
\includegraphics[scale=0.23]{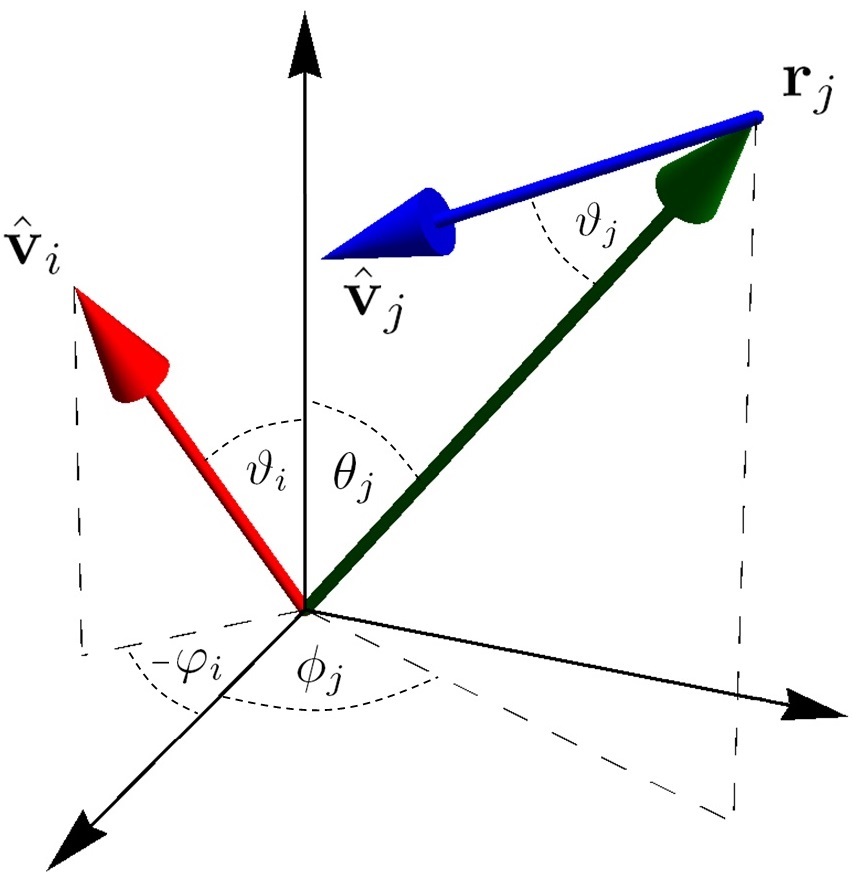}
\caption{Reference frame centred at $\rr_i$. Assuming that the gain functions are rotationally symmetric, then $G_j$ is a function of the angle between $\vv_j$ and $-\rr_j$ given by $\vartheta_j$, while $G_i$ a function of the angle between $\vv_i$ and $\rr_j$ given by $\chi=\arccos( \cos\theta_j \cos\vartheta_i + \cos(\phi_j - \varphi_i) \sin\theta_j \sin\vartheta_i )$. Note that when $\vartheta_i =0$, then $\chi=\theta_j$.}
\label{fig:cords}
\end{figure}

\section{Homogeneous Anisotropic Connectivity Mass \label{sec:funct}}

Since the system is homogeneous, we chose $\rr_i$ as the origin of the reference frame.
Furthermore, since the system is isotropic we can conveniently choose $\vv_i= (1,0,0)$ in spherical coordinates.
The position coordinates of node $j$ are given by $\rr_j = (r_j, \theta_j,\phi_j)$, while its orientation by $\vv_j=(1,\vartheta_j,\varphi_j)$ as shown in Fig.~\ref{fig:cords}.
Note that since the gain functions are rotationally symmetric, $\varphi_j$ is a free parameter.

We may now write the pair connectedness function explicitly as
\es{
H(r_j,\theta_j,\vartheta_{j})= \exp\pr{ -\frac{\beta r_j^{\eta}}{ G_{i}(\theta_j) G_{j}(\vartheta_j)  } }
.
\label{HH}}
Notice that $G_i$ is a function of the position of $j$ whilst $G_j$ is a function of the orientation of antenna $j$ as described in the caption of Fig.~\ref{fig:cords}.
Performing the $r_j$, $\phi_j$ and $\varphi_j$ integrals in \eqref{isoM} and simplifying we arrive at our first main result
\es{
&\mathsf{M} = \frac{1}{4\pi}\int \!\! \int_{0}^{2\pi} \!\! \int_{0}^{\pi} \!\! \int_{0}^{\infty} \!\! r_j^2 \sin \theta_j  
H(r_j,\theta_j,\vartheta_{j})
\dd r_j \dd \theta_j \textrm{d} \phi_j \dd \Omega_j
\\
&=  \frac{\pi\Gamma(3/\eta)}{ \eta \beta^{3/\eta}} \!
\pr{\!\int_{0}^{\pi} \!\!\!\! \sin\theta_j G_i(\theta_j)^{3/\eta} \dd \theta_j \!} \!\!
\pr{\!\int_{0}^{\pi} \!\!\!\! \sin\vartheta_j G_j(\vartheta_j)^{3/\eta} \dd \vartheta_j \!}
\label{masss}}
where $\Gamma(x)$ is the gamma function and the separation of the integrals is analogous to that obtained in the 2D case \cite{coonanisotrop}.
We notice that when $\eta=3$, the connectivity mass $\mathsf{M}$ is invariant\footnote{A similar observation was made for two dimensional domains in \cite{coonanisotrop} where the critical path loss was found to be $\eta=2$.} with respect to the specific radiation pattern due to the normalization condition in \eqref{norm}. 
Moreover, it follows from the structure of \eqref{isoM} that in a $d$-dimensional homogeneous domain e.g. $\V= \R^d$ the connectivity mass would involve a radial integral of $G^{d/\eta}$ such that $\mathsf{M}$ becomes invariant with respect to $G$ when $\eta=d$.
Therefore, we conclude that the ratio $d/\eta$ is a key system parameter whose importance will be highlighted in the following subsections.

Since the gain integrals in \eqref{masss} of $G_i$ and $G_j$ factor out nicely and are equivalent to each other, we define 
\es{
S_{\eta}[G]= \int_{0}^{\pi}  \sin\theta G(\theta)^{3/\eta}  \dd \theta 
\label{SG},}
and investigate its dependence with respect to the path exponent $\eta$, for simple but practical gain functions in order to identify which ones yield better (or worse) connectivity properties.

\subsection{Isotropic Radiation \label{sec:iso}}

As a benchmark for our theoretical analysis we set $G=1$ corresponding to isotropic radiation. 
In this case we have the following trivial result for the functional of interest
\es{
S_{\eta}[G] = \int_{0}^{\pi} \sin\theta  \textrm{d} \theta = 2
.\label{isot}}

\subsection{Wide-Angle Unidirectional Radiation \label{sec:patch}}

\begin{figure}[h]
\centering
\includegraphics[scale=0.2]{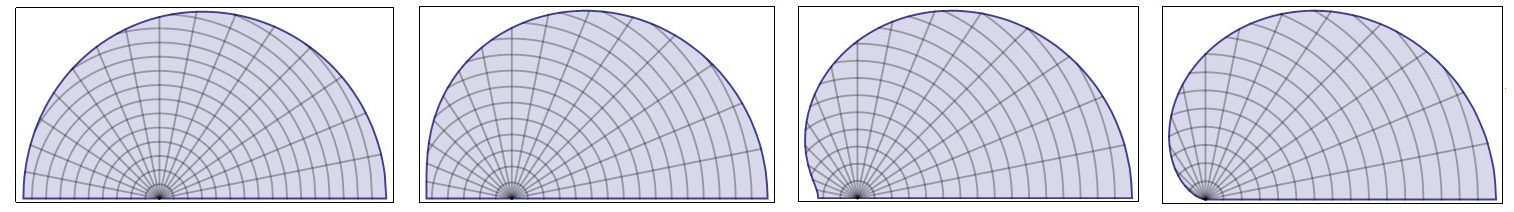}
\caption{The cardioid pattern for $\epsilon=1/4,1/2,3/4,$ and $1$, from left to right.}
\label{fig:cardiod}
\end{figure}

Wide-angle unidirectional radiation patterns are characteristic of microstrip antennas, also called patch antennas. 
Patch antennas are relatively cheap and easy to manufacture using modern printed-circuit technology. Moreover, they are mechanically robust and therefore are generally used in wireless communications where size, weigh, cost, performance, and ease of installation are often the main constraints \cite{balanis2012antenna}.
Their narrow bandwidth is being somewhat mitigated by the fact that many communications protocols are nowadays moving towards CDMA and TDMA techniques which use a single band.

Patch antenna gains typically have a single wide-angled major lobe with a number of small minor ones. 
We ignore the minor lobes and approximate the patch antenna radiation pattern by the cardioid function $G(\theta)=1+ \epsilon \cos\theta$, for $\theta\in(0,\pi)$ with $\epsilon\in[0,1]$.
The parameter $\epsilon$ measures the extent of deformation from the isotropic case as shown in Fig.~\ref{fig:cardiod} with $\epsilon=1$ corresponding to the most directional case.
To obtain the 3D radiation pattern, the gain profile of Fig.~\ref{fig:cardiod} is rotated about the $x$-axis, thus producing a surface of revolution.
For general $\epsilon$ we find
\es{
S_{\eta}[G]= \eta\frac{(1+\epsilon)^{1+3/\eta}-(1-\epsilon)^{1+3/\eta}}{\epsilon(\eta+3)}
\label{car}.}
Notice that when $\epsilon=0$ we recover \eqref{isot}, whilst when $\epsilon=1$ we have $S_{\eta}[G]= \frac{2^{1+3/\eta}}{1+3/\eta}$.

\subsection{Omnidirectional Radiation \label{sec:dipole}}

\begin{figure}[h]
\centering
\includegraphics[scale=0.2]{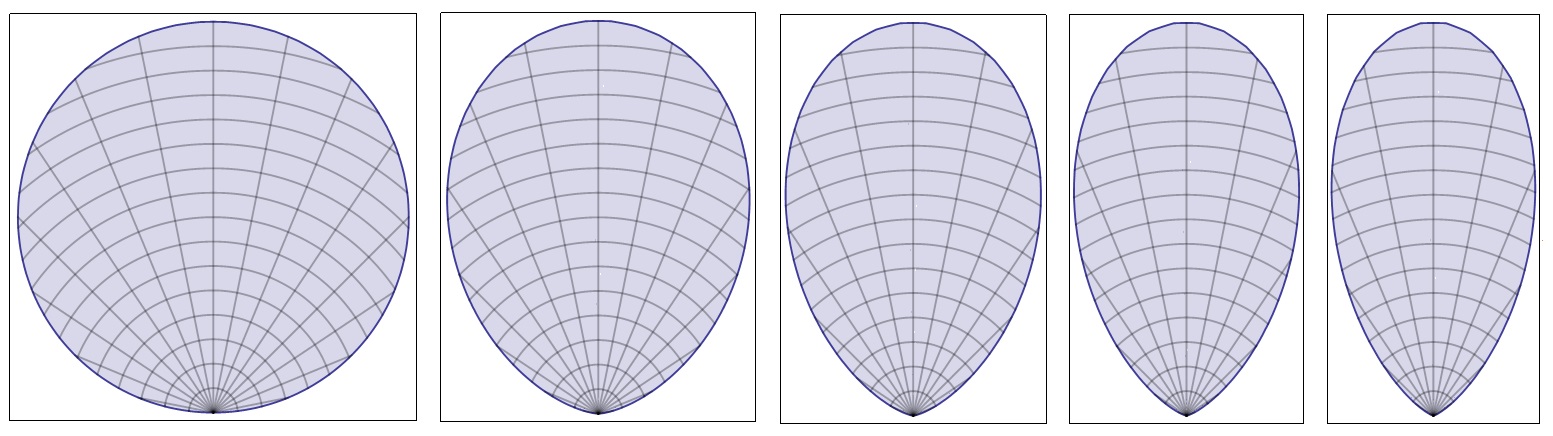}
\caption{The donought pattern for $m=1,\ldots ,5$, from left to right.}
\label{fig:donut}
\end{figure}

Omnidirectional radiation patterns are characteristic of dipole antennas, commonly used in wideband wireless applications.
Their radiation pattern is shaped like a doughnut and is symmetric about the axis of the dipole. 
In the simplest case of a half-wavelength dipole antenna we may approximate this pattern by $G(\theta)=\frac{2 \Gamma(\frac{3+m}{2})}{\sqrt{\pi}\Gamma(\frac{2+m}{2})} \sin^m \theta$, for $\theta\in(0,\pi)$ with $m>0$ \cite{balanis2012antenna}.
The parameter $m$ measures the directivity of the donought ring as shown in Fig.~\ref{fig:donut}.
For general $m$ we find
\es{
S_{\eta}[G]= \pr{\frac{2 \Gamma(\frac{3+m}{2})}{\sqrt{\pi}\Gamma(\frac{2+m}{2})}}^{3/\eta}
\frac{\sqrt{\pi} \Gamma(1+\frac{3m}{2\eta})}{\Gamma(\frac{3(m+\eta)}{2\eta})}
\label{don}.}

\subsection{Narrow-Angle Unidirectional Radiation \label{sec:endfire}}

\begin{figure}[h]
\centering
\includegraphics[scale=0.2]{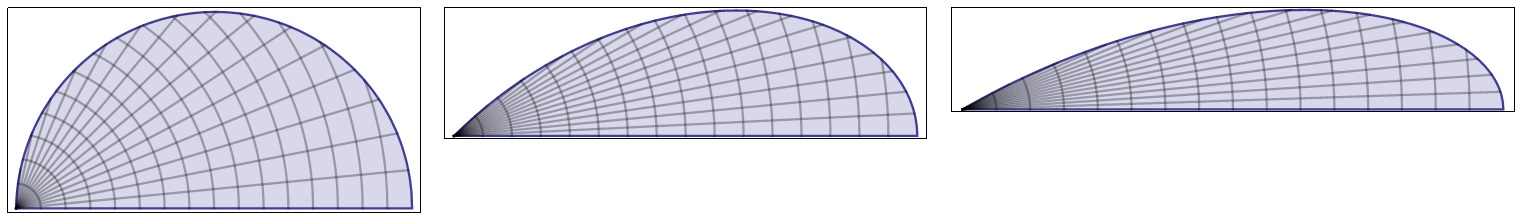}
\caption{The highly directional horn pattern for $\lambda=1,2,3$, from left to right.}
\label{fig:zep}
\end{figure}

Highly directional patterns are characteristic of horn antennas and can also be generated by end-fire arrays.
Typical applications include high powered satellite communications and radio telescopes, although beamforming techniques have recently attracted much attention in low-power wireless multihop networks \cite{bettstetter2005does,zhou2009connectivity}.
We approximate their radiation pattern by $G(\theta)=  (2(\lambda^2 -1)\cos\lambda\theta)/(\lambda \sin\frac{\pi}{2\lambda} -1)$ in the interval $\theta\in(0,\frac{\pi}{2\lambda})$ with for $\lambda \geq 1$ and $G=0$ elsewhere.
The parameter $\lambda$ measures the directivity of the beam as shown in Fig.~\ref{fig:zep}.
For the case of $\lambda=2$ we find
\es{
S_{\eta}[G]= \frac{\eta(6+6\sqrt{2})^{3/\eta}  }{2(3+\eta)} {}_2 F_1 (1,\frac{3}{2}+\frac{3}{2},2+\frac{3}{\eta},-1)
,
\label{zep}}
where $_2 F_1$ is the Gauss hypergeometric function.
Closed form expressions exist for other values of $\lambda$ but become increasingly complicated and do not offer further insight.

\begin{figure}[t]
\centering
\includegraphics[scale=0.185]{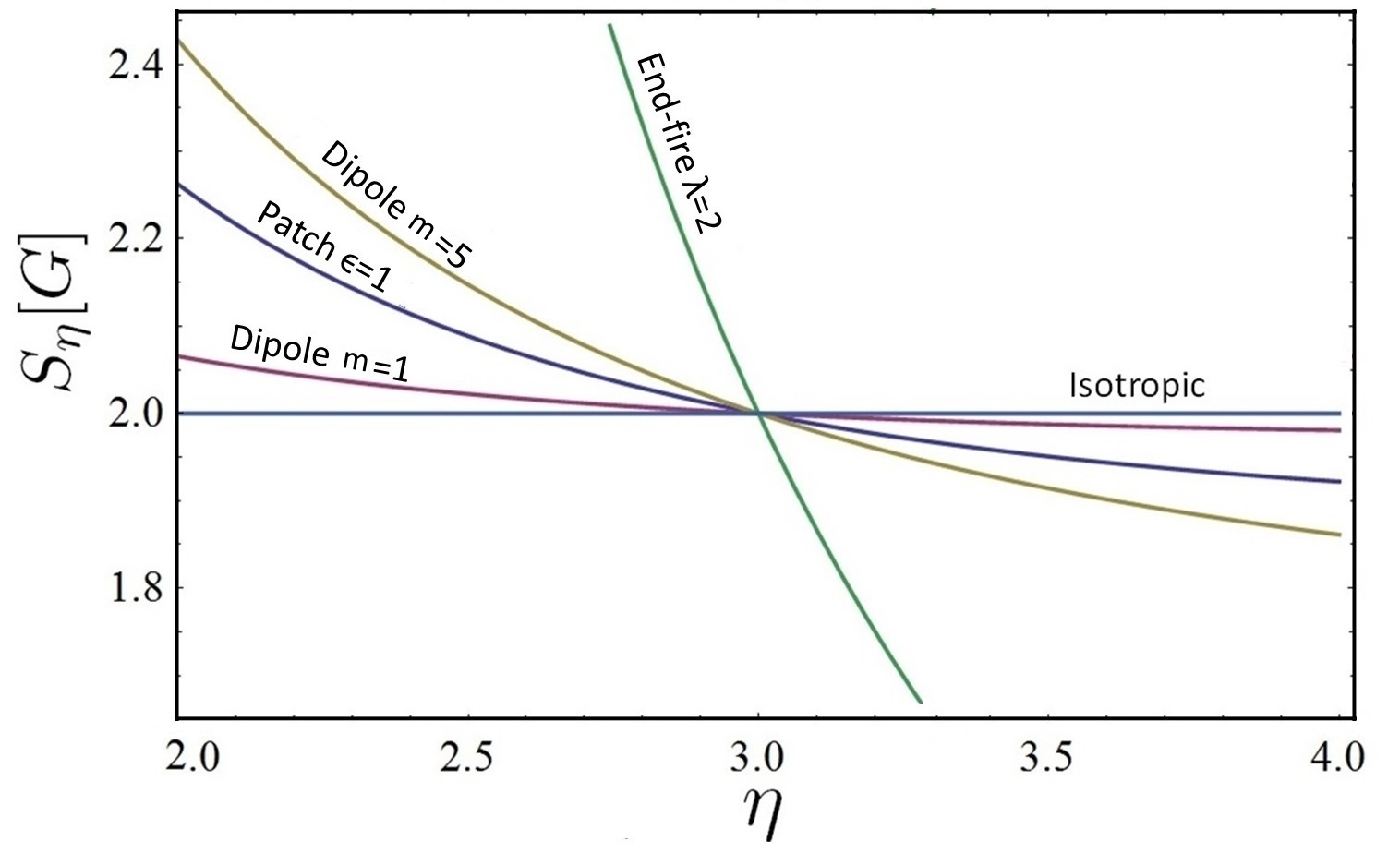}
\caption{Comparison of the functional $S_{\eta}[G]$ as a function of the path loss exponent $\eta$ for the various radiation patterns considered in this section. Directional radiation patterns are superior to the isotropic case only when $\eta<3$.}
\label{fig:compare}
\end{figure}

\subsection{Narrow-Angle Multi-directional Radiation \label{sec:endfire2}}

As a generalization to the above unidirectional radiation pattern we allow for the possibility of multiple, non-overlapping, highly directional, identical lobes (see the left panel of Fig.~\ref{fig:spikes} for an example with $n=6$ lobes). 
Although such exotic gain patterns are not often encountered in practice\footnote{Experimental realizations of multi-directional radiation patterns have reported substantial benefits and point towards successful application to large-scale wireless sensor networks \cite{giorgetti2007exploiting}.}, the following theoretical investigation presents an interesting exercise and can offer some useful design recommendations.
For a radiation pattern with $n>1$, the gain function is given by \eqref{multi} with each lobe described by $g_k(\theta^{(k)})=  (2(\lambda^2 -1)\cos\lambda\theta^{(k)})/(n\lambda \sin\frac{\pi}{2\lambda} -n) $ for $k=1,\ldots,n$, with $\theta^{(k)}\in(0,\frac{\pi}{2\lambda})$, and $\lambda \geq 1$.
For the case of $\lambda=2$ and general $n$ we find
\es{
S_{\eta}[G]&= \sum_{k=1}^{n} \int_{0}^{\frac{\pi}{2\lambda}} \sin (\theta^{(k)} ) g_k(\theta^{(k)})^{3/\eta} \dd \theta^{(k)} \\
&= \frac{n^{1-3/\eta}\eta(6+6\sqrt{2})^{3/\eta}  }{2(3+\eta)} {}_2 F_1 (1,\frac{3}{2}+\frac{3}{2},2+\frac{3}{\eta},-1)
,
\label{zep2}}
where we have assumed no overlapping lobes and thus considered each lobe's contribution to the integral individually.
Notice that for $\eta<3$, increasing the number of lobes $n$, \textit{ceteris paribus}, has the effect of decreasing $S_\eta[G]$. 
This is particularly interesting since it implies for example that at low path loss ($\eta=2$) and identical receive and transmit gains (i.e. $G_i = G_j$), doubling the number of lobes (normalized at constant total power \eqref{norm}) would result to halving the network mean node degree $\mu$.
Similarly, at high path loss ($\eta=6$), doubling the number of lobes doubles $\mu$.

\subsection{Single Sector Radiation \label{sec:sector}}

To simplify matters, we also consider a sectorized radiation model \cite{ramanathan2001performance,korakis2003mac} where $G(\theta)= f(\nu)=\textrm{const}>0$ for the interval $\theta\in(0,\nu \pi)$ for some $\nu \leq 1$ and $G=0$ elsewhere.
This would result in a conical radiation pattern ending in a spherical cap.
In complete analogy to the 2D case \cite{coonanisotrop}, by employing Lagrange multipliers in the calculus of variations we find that the constant gain function $G(\theta)=\csc^2 (\frac{\nu \pi}{2})$ yields the stationary path of $S_\eta [G]$.
For general $\nu$ we find
\es{
S_{\eta}[G]=2 \prr{\sin\pr{\frac{\nu \pi}{2}}}^{2-\frac{6}{\eta}}
,
\label{single}}
implying that the path defined by the isotropic radiation gain (i.e. when $\nu=1$) is a maximum of $S_{\eta}[G]$ for $\eta>3$,  and a minimum for $\eta<3$.
Therefore, we may conclude that isotropic radiation offers optimal connectivity properties when $\eta>3$ but the worst possible when $\eta<3$.

\begin{figure}[t]
\centering
\includegraphics[scale=0.18]{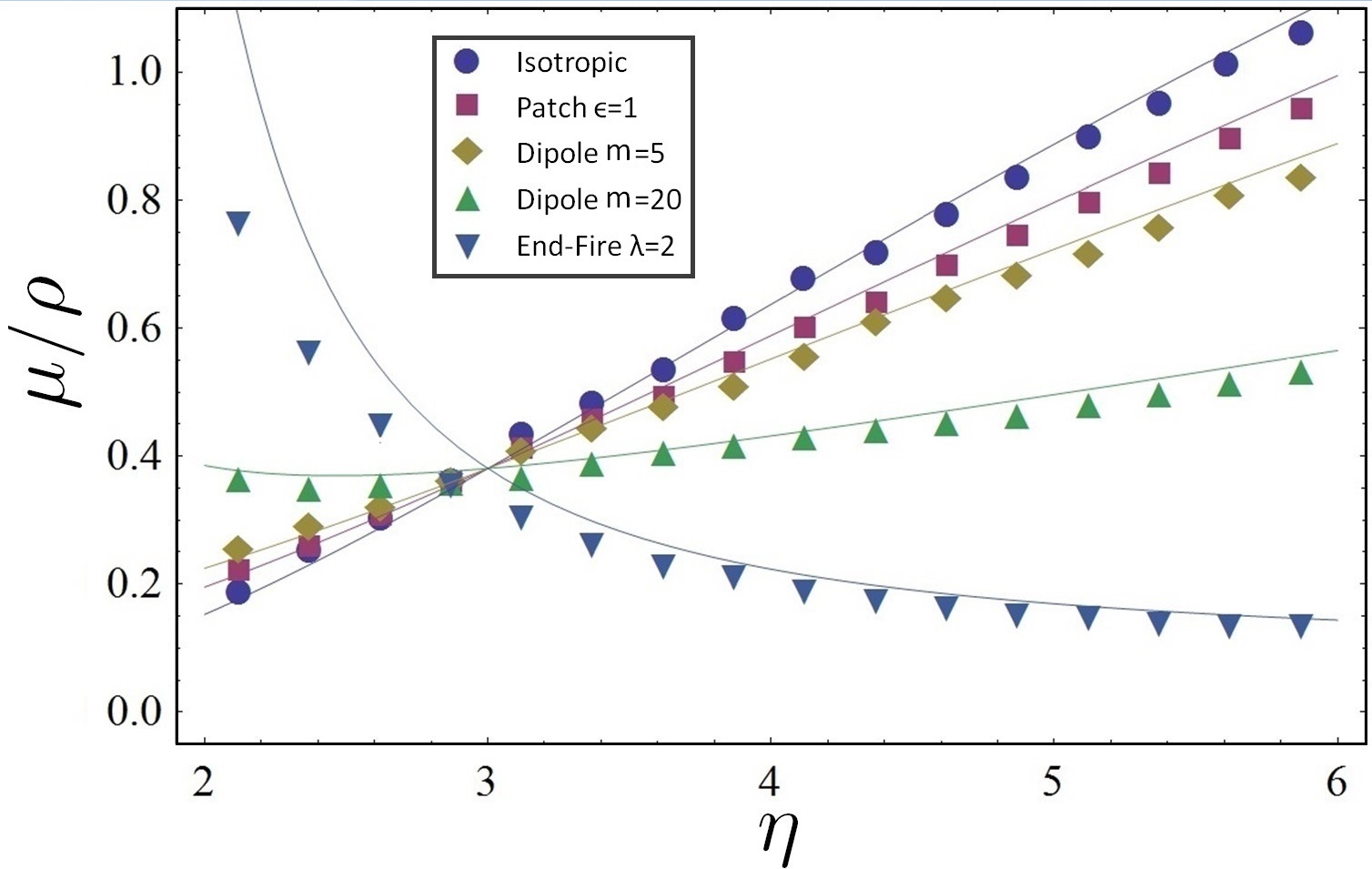}
\caption{Comparison of the computer simulated mean degree $\mu$ (showing using solid markers) and the theoretical prediction (curved line) for different antenna gains and for a range of path loss values $\eta\in[2,6]$. The simulation was run in a cube domain of side $L=10$, at a density $\rho=0.1$ and $\beta=10$.}
\label{fig:num}
\end{figure}

\subsection{Numerical Verification and Discussion}

Fig.~\ref{fig:compare} provides a qualitative comparison of $S_{\eta}[G]$, and thus $\mathsf{M}$, for the various gain functions considered above.
It is clear that directional radiation patterns can significantly improve network connectivity in homogeneous domains at low path loss $\eta<3$.
This observation is in good agreement with the numerical results in two dimensional networks of \cite{zhou2007analytical,zhou2009connectivity}, yet further highlights the importance of the ratio $d/\eta$ as well as its generality to any directional antenna gain profile or small-scale fading model\footnote{The ratio $d/\eta$ was shown to play an important role when considering diversity scaling laws for SIMO, MISO and MIMO systems \cite{coon2012connectivity}. Specifically, when $d/\eta>1$, one should expect progressive (i.e. superlinear) improvements in power and diversity scaling. It would therefore be interesting to investigate directivity and diversity in parallel from the point of view of the ratio $d/\eta$.}.
Significantly, equation \eqref{masss} suggests that homogeneous networks with antennas which have different receive and transmit gains, for example directional transmit but isotropic receive gains (often adopted to avoid antenna misalignments) are in fact in great disadvantage to directional-directional for $\eta<3$ and isotropic-isotropic for $\eta>3$.
This is a generic observation, independent of $i)$ the gain pattern details (e.g. minor lobes or sectorized approximation), $ii)$ the fading model used, and should therefore be contrasted against the multitude of related research works (see \cite{li2009asymptotic,banerjee2012self,ding2012directional} and references therein).

Fig.~\ref{fig:num} shows a comparison between theory and computer simulations. 
The observable of choice is the mean degree $\mu$ of a random network with randomly oriented antennas at different path loss values within the range of $\eta\in(2,6)$.
Note that in Fig.~\ref{fig:num} we divide $\mu$ by $\rho$ since for $V\gg1$, the mean degree increases linearly with the number of nodes in the network. 
This allows for a direct comparison with the theoretical prediction of the homogeneous connectivity mass $\mathsf{M}$.
The simulation confirms that at $\eta=3$ all curves meet and the mean degree is independent of the gain pattern details.
A good agreement is observed between theory and simulation,
although the theoretical curve appears to systematically overestimate that of the simulation data.
The reason for this is that the numerical simulation was performed in a finite cube domain (of side length $L=10$) therefore inducing boundary effects, which have thus far been ignored in the theoretical model under the assumption of a homogeneous system. 
Therefore, in the numerical simulations, nodes near the boundary may occasionally steer their main beam outside of the domain and are typically of lower degree.
While this phenomenon applies to all directional radiation patterns, it is most significant for patterns whose gain is concentrated over a narrow solid angle.
Indeed, the curve for the highly directional end-fire array radiation pattern is noticeably above that of the data at low path loss $\eta<3$.
We will elaborate on this further in Sec. \ref{sec:BE}.



\section{Directivity Scaling \label{sec:DS}}

Expressions \eqref{don}, \eqref{zep} and \eqref{single} have highly directional limits, for $m\to\infty$,
$\lambda\to\infty$, and $\nu\to 0$ respectively, in which the gain pattern and derived quantities scale.
Physically we see that if the gain $G$ is concentrated on a small solid angle $\omega$,
hence (due to normalisation) having values of order $\omega^{-1}$, the integral
$S_\eta[G]$ will scale as $\omega^{1-3/\eta}$. 
We can see this in more detail for each of the two models.
Define $\omega$ to be the solid angle over which $G$ takes at least half its maximum value.
Then for the dipole case of $G(\theta)=\frac{2 \Gamma(\frac{3+m}{2})}{\sqrt{\pi}\Gamma(\frac{2+m}{2})} \sin^m \theta$,
we require that $\sin^m\theta\geq 1/2$, which gives
a small interval (to leading order in $m^{-1}$): $|\theta-\pi/2|\leq \sqrt{2\ln 2/m}$. 
Multiplying the width of this interval $2\sqrt{2\ln 2/m}$ by the length of the equator $2\pi$ gives
\es{
\omega=\sqrt{\frac{32\pi^2\ln 2}{m}},
\label{omega1}}
for $m\to\infty$. Applying the asymptotic formula for the ratio of gamma functions in \eqref{don} 
\es{
\frac{\Gamma(z+a)}{\Gamma(z+b)}\sim z^{a-b},
}
(see \cite{NIST:DLMF} 5.11.12) for $z\to\infty$ and comparing with \eqref{omega1} we find that
\es{
S_\eta[G]\sim C_1(\eta)\omega^{1-3/\eta}
,}
for an explicit (but rather unilluminating) function $C_1(\eta)$.

For the highly directional radiation pattern $G(\theta)=  (2(\lambda^2 -1)\cos\lambda\theta)/(\lambda \sin\frac{\pi}{2\lambda} -1)$ for $\theta\in(0,\frac{\pi}{2\lambda})$,
we use the same definition of $\omega$, this time finding
$\theta\leq \pi/(3\lambda)$ and hence
\es{
\omega=4\pi \sin^2 \pr{\frac{\pi}{2\lambda}}\sim \frac{\pi^3}{9\lambda^2}
.\label{omega}}
Making a change of variable $t=\lambda\theta$ in the integral of interest, we find
\es{
S_\eta[G]=\left(\frac{2(\lambda^2-1)}{\lambda\sin(\pi/2\lambda)-1}\right)^{3/\eta}
\int_0^{\pi/2}\frac{\sin(t/\lambda)\cos^{3/\eta}t}{\lambda} \dd t
.}
Taking the limit $\lambda\to\infty$, the sines take their forms for small argument, and so
\es{
S_\eta[G]&\sim\left(\frac{2}{\pi/2-1}\right)^{3/\eta}\lambda^{6/\eta-2}\int_0^{\pi/2}
t\cos^{3/\eta}t \dd t \\
&\sim C_2(\eta)\omega^{1-3/\eta}
,}
where now $C_2(\eta)$ involves a non-elementary integral for most values of $\eta$.

Finally, for the single sector radiation pattern we have that in the directional limit of $\nu\to 0$, equation \eqref{single} becomes
\es{
S_{\eta}[G]&\sim 2\pr{\frac{\nu \pi}{2}}^{2-\frac{6}{\eta}}\\
&\sim 2^{\frac{3}{\eta}} \pr{\frac{\omega}{2\pi}}^{1-\frac{3}{\eta}}
,}
since $\omega= 2\pi (1-\cos \nu \pi) = \pi^3 \nu^2 + \mathcal{O}(\nu^4)$.

Significantly, we find that when $G_i = G_j$, in either of the above three cases cases the connectivity mass scales as
$\mathsf{M}\sim C_3(\eta)\omega^{2-6/\eta}$. 
Thus when $\eta<3$, scaling the density as $\rho\sim\omega^{6/\eta-2}$ will keep the exponent $\rho \mathsf{M}$ in \eqref{isoP} constant but result in a decrease in $N$, and hence lead to an increase in the probability of full connectivity.
For example, when $\eta=2$, this suggests that by making a directional beam of half the solid angle, we may reduce the number of nodes by a significant factor of $2$ without affecting $\mathsf{M}$ or any of the associated network connectivity properties described in Sec. \ref{sec:NC}.
We conclude this section by noting that it is reasonable to expect our results to generalize to general dimension $d$ such that $\mathsf{M}\sim \omega^{2-2d/\eta}$.

\begin{figure}[t]
\centering
\includegraphics[scale=0.15]{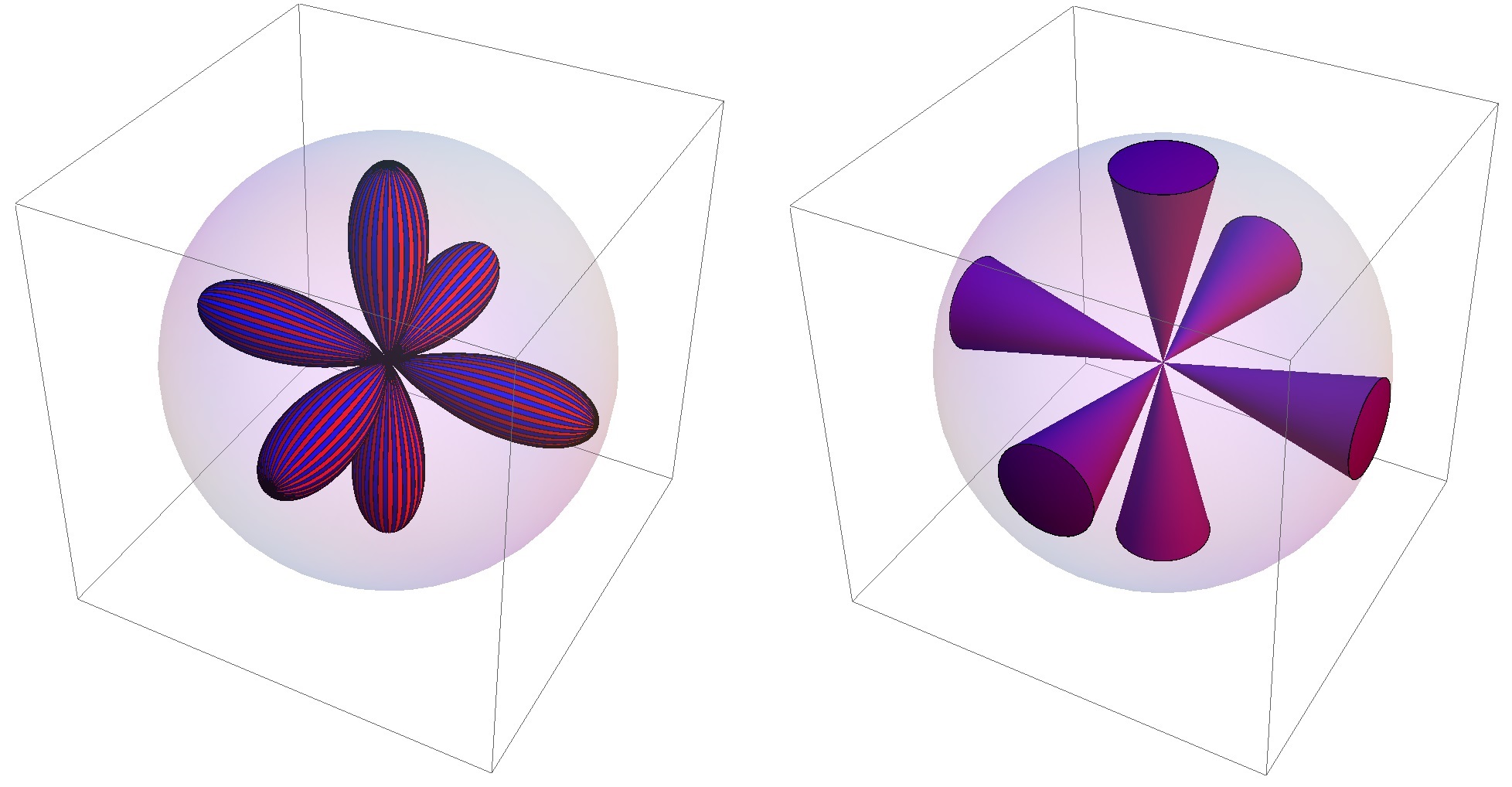}
\caption{An example of a narrow-angled multi-directional radiation pattern with $n=6$ evenly spaced non-overlapping lobes is shown on the left with its corresponding multi-sectorized approximation on the right.}
\label{fig:spikes}
\end{figure}


\section{Inhomogeneous Anisotropic Connectivity Mass and Boundary Effects \label{sec:BE}}

Random networks confined within a bounded domain $\V\subset \R^3$ are no longer homogeneous nor isotropic. 
As a result, boundary effects can have a significant impact on the connectivity properties of such networks \cite{khalid2013connectivity,coon2012impact,fabbri2008multi,fan2007geometrical,guo2013outage}.
The main reason behind this is that nodes situated near the boundary have a higher probability of being isolated (i.e. of degree $0$) than nodes in the bulk component of $\V$. 
This feature is quantified by \eqref{Pfciso} in the isotropic case, and further intensified in the case of anisotropic radiation patterns where nodes near the boundary may steer their connectivity beam(s) outside the network domain and hence increase their isolation likelihood.
Therefore, unlike in homogeneous systems where directivity significantly improves network connectivity at low path loss, in inhomogeneous systems, directional radiation patterns present us with some serious drawbacks.

In order to understand the negative effects of directional radiation patterns, we examine the high density asymptotic behaviour of $P_{fc}$. 
From \eqref{Pfc}, we expect this behaviour to stem from the extreme situation where $\rr_i$ is situated exactly at the sharpest corner of $\V$.
For simplicity we will consider a right angled corner\footnote{Without this assumption all the results generalize but exposition and notations become a little more cumbersome.} and naturally choose this as the origin of the reference frame, with the corner edges aligned with the positive coordinate axes.
We also restrict the current discussion to rotationally symmetric gain patterns and thus exclude the multi-directional case.
Note that the system is not isotropic and so $G_{i}$ is a function of the angle between $\vv_i$ and $\rr_j$ given by $\chi=\arccos( \cos\theta_j \cos\vartheta_i + \cos(\phi_j - \varphi_i) \sin\theta_j \sin\vartheta_i )$ as illustrated in Fig.~\ref{fig:cords}.
We therefore have that 
\es{
P_{fc} \approx 1 - \frac{\rho}{4\pi} \int e^{-\rho \mathsf{M}_{C}} \dd \Omega_i ,
}
where $\mathsf{M}_C$ is the connectivity mass associated with the right angled corner and is given by averaging $H_{ij}$ over all possible positions and antenna orientations of node $j$
\es{
\mathsf{M}_C &= \! \frac{1}{4\pi} \!\!  \int \!\! \int_{0}^{\frac{\pi}{2}} \!\! \int_{0}^{\frac{\pi}{2}} \!\! \int_{0}^{\infty} \!\! r_j^2 \sin\theta_j e^{-\frac{\beta r_j^\eta}{G_i(\chi) G_j(\vartheta_j)}} \dd r_j \dd \theta_j \dd \phi_j  \dd \Omega_j \\
&= \! \frac{\Gamma(3/\eta)}{4 \pi \eta \beta^{3/\eta}} \!\! \int  \!\! \int_{0}^{\frac{\pi}{2}} \!\! \int_{0}^{\frac{\pi}{2}}\!\! \sin\theta_j  (G_i(\chi) G_j(\vartheta_j))^{3/\eta} \dd \theta_j \dd \phi_j \dd \Omega_j \\
&= \! \frac{\Gamma(3/\eta)}{2\eta \beta^{3/\eta}}
  \pr{  \int_{0}^{\frac{\pi}{2}}  \!\! \int_{0}^{\frac{\pi}{2}} \sin\theta_j   G_i(\chi)^{3/\eta} \dd \theta_j \dd \phi_j } S_{\eta}[G_j]
,\label{MC}}
where we have extended the radial integral in the first line of \eqref{MC} to infinity (even though $\V$ is finite) since $H_{ij}$ is decaying exponentially - a reasonable approximation if the connectivity range is much smaller than the size of the system.
Note that unlike the homogeneous case where the connectivity mass was invariant to the gain details at $\eta=3$, here, $\mathsf{M}_C$ depends strongly on $G_i$ through the antenna orientation of the cornered node $i$ for all values of $\eta$.
Consequently, in order to make further progress, we now focus at the asymptotic exponential decay of $P_{fc}$ given by the minimum of $\mathsf{M}_{C}$ over the orientation of antenna $i$ which we henceforth denote as $\M$.
Therefore, we must seek the minimum value of the integral
\es{
I_{G_i}(\vartheta_i,\varphi_i) = \int_{0}^{\frac{\pi}{2}}  \!\! \int_{0}^{\frac{\pi}{2}} \sin\theta_j   G_i(\chi)^{3/\eta} \dd \theta_j \dd \phi_j
\label{Imin}}
with respect to the orientation vector $\vv_i$.
We do this for the four simple gain functions described in the previous section.

\subsection{Isotropic Radiation}

In this case $G=1$ and we obtain the trivial result that $I_{G_i} =\pi/2$ and so $\M= \mathsf{M}_C = \frac{\Gamma(3/\eta) \pi}{2\eta \beta^{3/\eta}}$. 
More generally, for corner of solid angle $\omega_C$ we would have $\M= \mathsf{M}_C= \frac{\Gamma(3/\eta)}{\eta \beta^{3/\eta}} \omega_C$.

\subsection{Wide-Angle Unidirectional Radiation}

In this case, $G(\theta)= 1+\epsilon \cos\theta$, which leads to a global minimum in \eqref{Imin} at $(\vartheta_i,\varphi_i)=(\pi-\theta_j,\phi_j+\pi)$, i.e. when $\rr_i = -c \vv_i$ for any $c>0$ which can only occur if $\vv_i$ is pointing outside of $\V$.
Such a configuration amounts to
\es{
I_{G_{i}}(\pi-\theta_j,\phi_j+\pi)=\frac{\pi}{2}(1-\epsilon)^{3/\eta}
.}
Note that when $\epsilon=1$, we have that $I_{G_{i}}=0$ which is indicative of a blind spot, meaning that any node directly behind node $i$ finds it impossible to connect with it.
However such a configuration is highly unlikely as it is of zero probability measure.


\subsection{Omnidirectional Radiation}

In this case, $G(\theta)=\frac{2 \Gamma(\frac{3+m}{2})}{\sqrt{\pi}\Gamma(\frac{2+m}{2})} \sin^m \theta$, which leads to two global minima in \eqref{Imin} at $(\vartheta_i,\varphi_i)=(\pi-\theta_j,\phi_j+\pi)$ and $(\vartheta_i,\varphi_i)=(\theta_j,\phi_j)$, giving $I_{G_i}=0$ in both cases.
This is expected due to the zero gain in both $\pm\vv_i$ directions, however as with the previous case, such a configuration is highly unlikely.

\subsection{Narrow-Angled Unidirectional Radiation}

In this case, $G(\theta)=  (2(\lambda^2 -1)\cos\lambda\theta)/(\lambda \sin\frac{\pi}{2\lambda} -1)$, which leads to $I=0$ for a range of orientations independent of $(\theta_j,\phi_j)$, i.e. whenever the entire lobe is oriented outside of the domain $\V$.
Therefore, the blind-spot phenomenon is much more likely.
For this reason, we conclude that while highly directional antennas improve connectivity in homogeneous domains at low path loss, they are in some disadvantage to wide-angle unidirectional or omnidirectional antennas in inhomogeneous domains.


\section{Inhomogeneous Connectivity Mass for Multi-directional Radiation \label{sec:multi}}

Networks operating in homogeneous environments at low path loss can improve their connectivity by using highly directional antennas.
Networks operating in inhomogeneous domains however suffer from boundary effects (especially near sharp corners) where nodes with unidirectional radiation patterns (and especially narrow-angle ones) may steer their main beam outside the domain thus suffering from blind-spots.
In the absence of any \textit{a posteriori} knowledge or control over antenna orientations (e.g. beamstearing capabilities) it is therefore desirable to identify ways of mitigating blind-spots and achieve a high connectivity mass both near and away from the domain boundary. 

One possible approach to the above stated problem is to consider multi-directional patterns where the gain is concentrated on $n \geq 2$ \textit{evenly spaced} lobes.
Similar radiation patterns have been experimentally realized in \cite{giorgetti2007exploiting} using a number of patch antennas.
Since we are interested in the performance of $G$ in low path loss we now lift the assumption that the connectivity range is much smaller than the size of the system\footnote{This assumption was used to extend the radial integral of \eqref{MC} to infinity.}.

\subsection{Two Dimensional Case}

In order to aid in the discussion of the impact of boundaries in 3D multi-directional radiation patterns, we first discuss the 2D case.
In two dimensions, distributing $n\geq 2$ points evenly on the unit circle is a trivial problem with $\varTheta=\{\vv^{(k)}= (1, 2\pi k/n + x), \textrm{ for } k=1,\ldots,n \textrm{ and } x\in[0,2\pi/n)\}$.  
Each lobe has $g_k(\theta^{(k)})= \lambda \pi \cos \lambda \theta$ for $\theta\in(-\frac{\pi}{2\lambda},\frac{\pi}{2\lambda})$ and $\lambda\geq 1$, and the total gain profile $G$ is defined by \eqref{multi}.
Following the discussion in Sec. \ref{sec:DS}, we may simplify the multi-directional gain function for $\lambda\gg1$ by considering a multi-sectorized radiation model where each 2D lobe is approximated by a sector of angular width $\omega=2\pi/(3\lambda)$ and gain $g_k(\theta^{(k)})=3\lambda/n$ for $\theta^{(k)}\in(-\frac{\pi}{3\lambda},\frac{\pi}{3\lambda})$ measured from $\vv^{(k)}$ and $0$ otherwise, such that the total power is normalized by
\es{
\int_{0}^{2\pi} G(\theta)\dd\theta=  \sum_{k=1}^{n} \int_{-\frac{\pi}{3\lambda}}^{\frac{\pi}{3\lambda}} g_k(\theta^{(k)}) \dd \theta = 2\pi
.}
Also, to avoid overlap between lobes we require that $\lambda\geq 3n$.

\begin{figure}[t]
\centering
\includegraphics[scale=0.245]{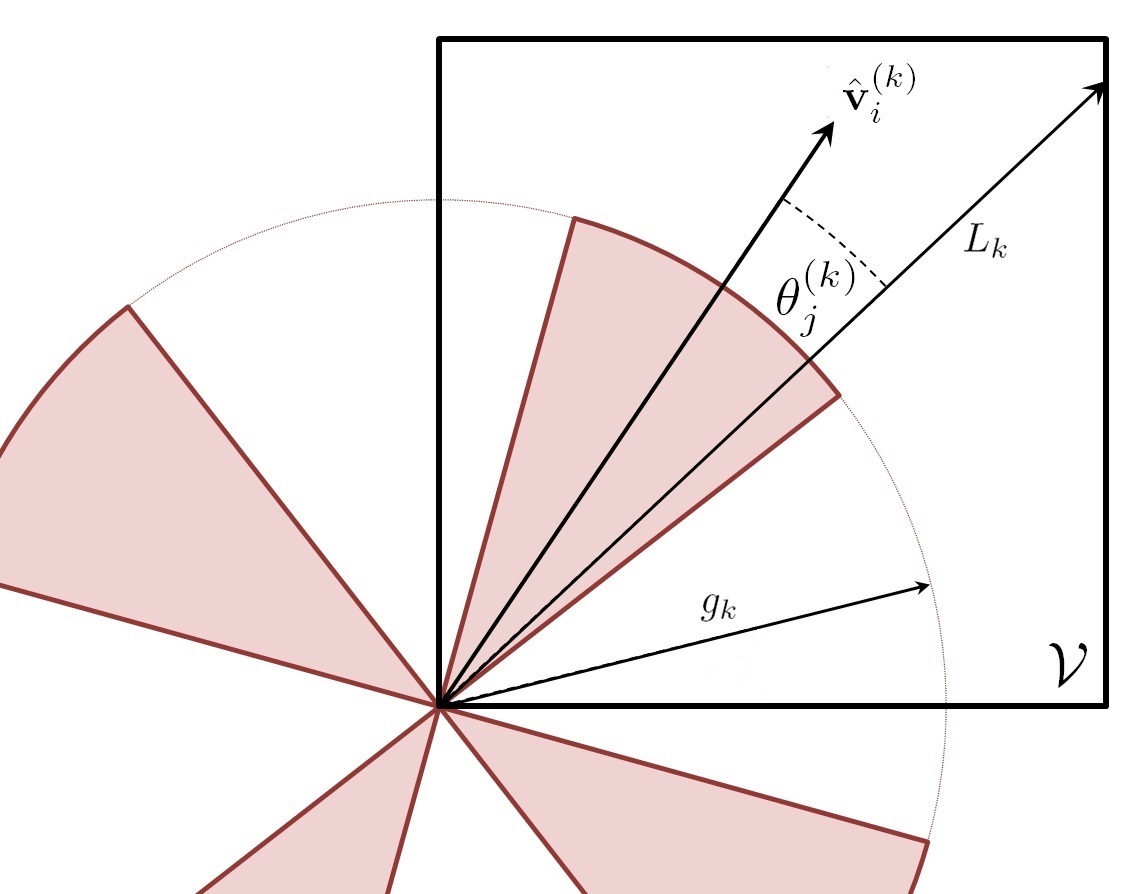}
\caption{Schematic of a multi-directional gain pattern ($n=4$) of a node situated at one of the corners of a square domain.}
\label{fig:2d}
\end{figure}

Having defined $G$, we now examine $\M=\min_{x} \mathsf{M}_C$ for a right angled corner of a square domain $\V\subset \R^2$.
The contribution to $\mathsf{M}_C$ from a single sector of antenna $i$ is given by
\es{
\frac{n}{2\pi}\!\int_{-\frac{\pi}{3\lambda}}^{\frac{\pi}{3\lambda}} \! \int_{-\frac{\pi}{3\lambda}}^{\frac{\pi}{3\lambda}} \! \int_{0}^{L_k} \!  r_j \exp\pr{\!\frac{ -\beta r_j^\eta}{g_k(\theta^{(k)})^2}\!}
\dd r_j \dd \theta_j^{(k)} \dd \vartheta_j ,
}
where the factor of $n$ in the front is due to the $n$ lobes of antenna $j$, and $L_k$ is the radial distance from the corner to the adjacent boundary of $\V$ in the direction $\theta_j^{(k)}$ measured from $\vv_i^{(k)}$, as illustrated for $n=4$ in Fig.~\ref{fig:2d}.
Performing the $\dd r_j$ integral and summing over $k$ we obtain
\es{
\mathsf{M}_C (n)= \frac{n g^{4/\eta}}{2\pi \eta \beta^{2/\eta}}\!\sum_{k=1}^{n}\! \int_{-\frac{\pi}{3\lambda}}^{\frac{\pi}{3\lambda}} \! \int_{-\frac{\pi}{3\lambda}}^{\frac{\pi}{3\lambda}} \!   \gamma\!\pr{\!\frac{2}{\eta},\frac{\beta L_k^{\eta}}{g^2}} 
 \dd \theta_j^{(k)} \dd \vartheta_j ,
}
where $\gamma(s,x)$ is the lower incomplete gamma function.
Note that we have dropped the subscript $k$ from $g$ as all lobes are identical.
For highly directional sectors ($\lambda\gg 1$), we may approximate $L_k$ by $\hat{L}_k$ given by the length of the vector $\vv_i^{(k)}$ projected onto the adjacent boundary of $\V$.
Notice that $\hat{L}_k$ is zero if the lobe is pointing outside the domain. 
After some simplifications we finally to arrive at
\es{
\M (n) \approx \min_{x} \prr{\frac{2\pi g^{4/\eta-2}}{ n \eta  \beta^{2/\eta}} \!\sum_{k=1}^{n}   \gamma\!\pr{\!\frac{2}{\eta},\frac{\beta \hat{L}_k^{\eta}}{g^2}} },
\label{2d}}
indicating that the finite size effect of truncating radial integration at $\hat{L}_k$ is of variable importance at different path loss.
Equation \eqref{2d} is difficult to calculate analytically, but straightforward numerically using a fine grid of values for $x\in[0,2\pi/n)$.
We now turn to the full problem in three dimensions.



\subsection{Three Dimensional Case}

In three dimensions, there are a number of ways of arranging $n> 2$ points evenly on the unit sphere.
One way is through the \textit{Thomson problem} (proposed in 1904 by J.J. Thomson, see \cite{wales2006structure} and references therein) concerning the minimum energy configuration of $n$ electrons confined on the surface of a sphere which repel each other with a Coulomb force.
Other ways involve packing and covering problems called the Tammes and Fejes T\'{o}th type problems respectively. 
We will adopt the Thomson interpretation for its connection with spherical molecule symmetries.

An analytic description of the $n$-point coordinate configuration is impossible. However, computer programs have generated them to very good accuracy and have also identified their symmetry types for very large values of $n$. 
One can try to imagine such configurations as the vertices of a polyhedron whose $2(n-2)$ faces are \textit{almost} equilateral triangles\footnote{For $n=4,6,$ and $12$, the triangles are perfect equilaterals and so the polyhedrons formed are the regular tetrahedron, octahedron, and icosahedron respectively.}.
An extensive table with the minimal energy, group symmetry, dual polyhedron and Cartesian coordinates of the $n$ vertices can be found online at \cite{webapp}.

We consider a multi-sectorized radiation model where each lobe is approximated by a cone ending in a spherical cap of radius $g_k=\frac{1}{n}\csc^2(\frac{\pi}{6\lambda})$, each of solid angle $\omega= 4\pi \sin^2 (\frac{\pi}{2\lambda})$ as in \eqref{omega}, thus satisfying the normalization condition \eqref{norm}.
Fig.~\ref{fig:spikes} shows an example multi-sectorized radiation model with $n=6$.
Notice that the resulting multi-directional gain pattern is not rotationally symmetric and so in general we have $\varTheta=\{ \vv^{(k)}=(1,\vartheta^{(k)},\varphi^{(k)}), \textrm{ for } k=1,\ldots,n \}$.
Finally, using $\sqrt{4\pi/n}$ as a rough estimate of the typical angular distance between neighbouring lobes we require $\lambda\geq \sqrt{n \pi}/3$ to avoid lobe overlapping.

\begin{figure}[t]
\centering
\includegraphics[scale=0.25]{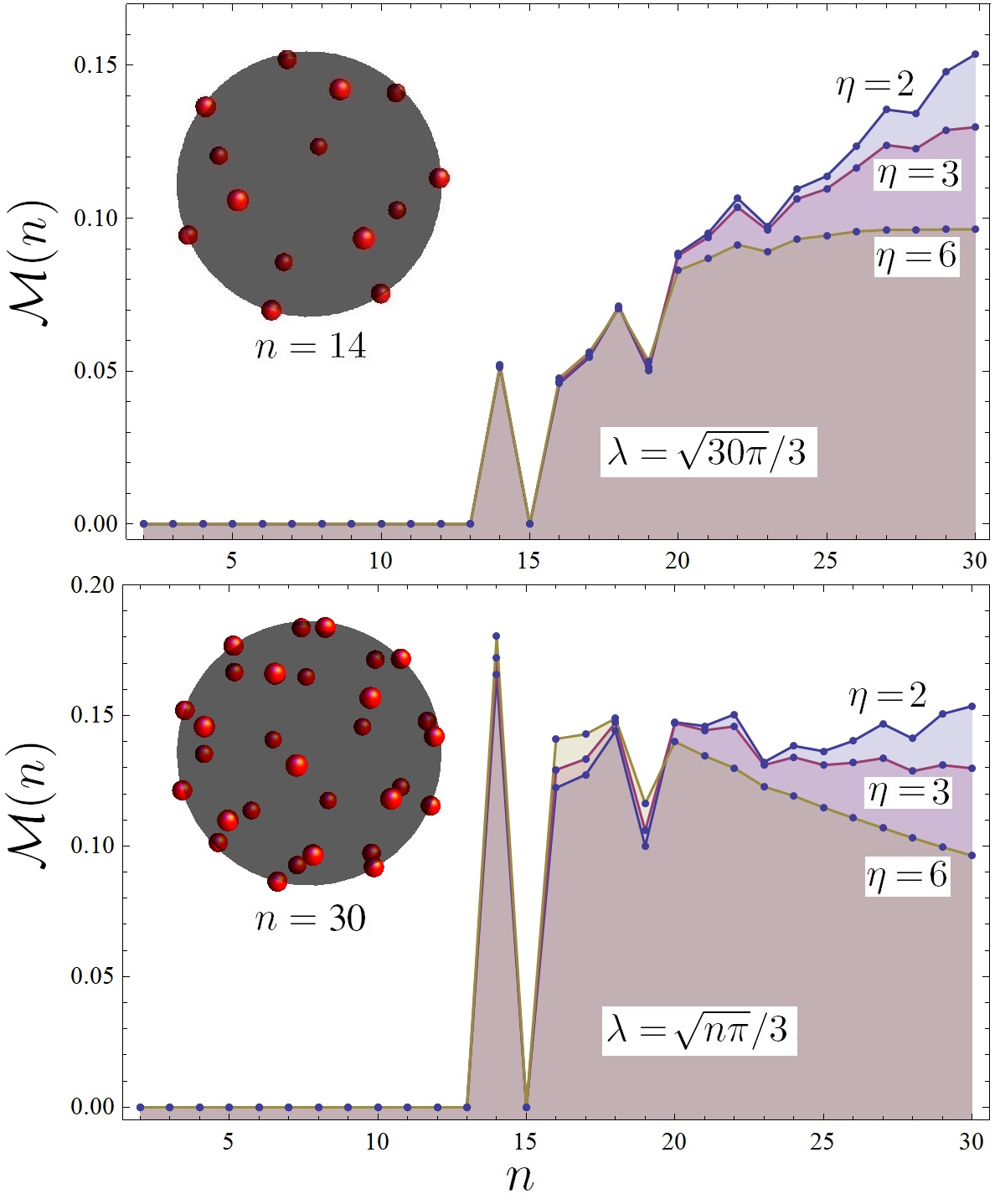}
\caption{Numerical calculation of the minimum connectivity mass \eqref{M3} due to a multi-directional antenna situated at right angled corner for $\eta=2,3,$ and $6$. 
The $n$ evenly spaced vertex coordinates were taken from \cite{webapp}, and the minimum is taken over all possible orientations $\varTheta_i$ using rotation matrices and a finite grid of Euler angles. 
The top panel keeps the lobe widths constant while the lower one shrinks them for increasing $n$ by scaling $\lambda$ by $\sqrt{n\pi}/3$.
The insets show the vertex positions for $n=14$ and $n=30$.}
\label{fig:minM}
\end{figure}

Having defined $G$, we now examine the connectivity mass $\mathsf{M}_C$ associated with a right angled corner of a cube domain $\V\subset \R^3$.
Taking the same approach as in the two dimensional case, we calculate
\es{
\mathsf{M}_C ( n)\! &= \!\frac{n}{4\pi}\sum_{k=1}^{n} \int \!\!\!  \int_{0}^{2\pi}  \!\!\!  \int_{0}^{\frac{\pi}{3\lambda}} \!\!\! \int_{0}^{L_k}
\!\!\!   
r_j^{2} \sin\theta_j e^{-\beta\frac{r_j^{\eta}}{g^2} }
 \dd r_j \dd \theta_j  \dd \phi_j \dd \Omega_j \\
&=
\! \frac{n g^{6/\eta}}{4\pi \eta \beta^{3/\eta}}\!\sum_{k=1}^{n} \! \int \!\!\! \int_{0}^{2\pi} \!\!\! \int_{0}^{\frac{\pi}{3\lambda}}
\!\!\! \gamma\!\pr{\!\frac{3}{\eta},\frac{\beta L_k^{\eta}}{g^2}}\!\sin\theta_j \dd \theta_j \dd \phi_j  \dd \Omega_j \\
&=\! \frac{ g^{6/\eta-1}}{ \eta \beta^{3/\eta}}\!\sum_{k=1}^{n} \! \int_{0}^{2\pi} \!\!\! \int_{0}^{\frac{\pi}{3\lambda}}
\!\!\! \gamma\!\pr{\!\frac{3}{\eta},\frac{\beta L_k^{\eta}}{g^2}}\!\sin\theta_j \dd \theta_j \dd \phi_j ,
\label{lobe}}
where we have restricted the $\dd \Omega_j$ integral over $\vartheta_j \in(0,\frac{\pi}{3\lambda})$ and $\varphi_j \in(0,2\pi)$ where the gain is non-zero.
Approximating $L_k$ by $\hat{L}_k$ as in the 2D case, we finally arrive at
\es{
\mathsf{M}_C (n) \approx \frac{ 4\pi g^{6/\eta-2}}{ n\eta \beta^{3/\eta}} \sum_{k=1}^{n}   \gamma \pr{\frac{3}{\eta},\frac{\beta \hat{L}_k^{\eta}}{g^2}} 
,
\label{lobe2}}
and thus obtain $\M$ by finding the minimum of $\mathsf{M}_{C}$ over all possible antenna orientations
\es{
\M ( n ) = \min_{\varTheta_{i}} \mathsf{M}_C (n)
.
\label{M3}} 

This is difficult to calculate analytically, but straightforward numerically, using rotation matrices and a fine grid of Euler angles.
The result is shown in Fig.~\ref{fig:minM} for the range $n\in[2,30]$ using a cube domain of side $L=1$, $\beta=1$, $\lambda=\sqrt{30\pi}/3$ in the top panel and $\lambda=\sqrt{n\pi}/3$ in the lower one. 
For comparison, all simulations were preformed for path loss values of $\eta=2,3,$ and $6$.
We observe that for $n\leq 13$, there always exist at least one orientation $\varTheta_i$ such that the pattern does not cover the cubic corner and therefore $\M=0$, i.e., the multi-directional gain has blind spots.
Interestingly, the case of $n=14$ (corresponding to a polyhedron called a ``gyroelongated hexagonal bipyramid" also shown in the inset of the top panel of Fig.~\ref{fig:minM}) covers such corners whilst that of $n=15$ does not. 
For larger values of $n\geq16$, blind spots are always covered, i.e., $\M>0$. For constant lobe widths characterised by $\lambda=\sqrt{30\pi}/3$, the minimum connectivity mass $\M(n)$ increases with $n$ modulo small fluctuations with better performance at lower path loss.
When the lobe widths are scaled by $\lambda=\sqrt{n\pi}/3$, the minimum connectivity mass is approximately constant at $\M\approx 0.15$ for $n\geq16$ (and $n=14$) when $\eta=2$, but decreases steadily for $\eta=3$, and $6$.
For comparison, we point out that isotropic radiation would give $\mathsf{M}_C\approx 0.416, 0.427,$ and $0.446$ for $\eta=2,3,$ and $6$ respectively, which is significantly higher than that observed for the multi-directional radiation patterns investigated in Fig.~\ref{fig:minM}. We stress however that Fig.~\ref{fig:minM} shows the minimum of $\mathsf{M}_C(n)$ over all possible orientations $\varTheta_i$ and further recall the benefits of directional patterns in homogeneous systems at low path loss as discussed in Sec. \ref{sec:funct}.

The above results hint towards an interesting generalization for arbitrarily shaped three dimensional domains.
Since at low path loss, increasing the number of lobes and scaling their widths by $\lambda=\sqrt{n\pi}/3$ improves $\M$ while also covering any corner, we propose as an optimal (yet unrealistic) limit a radiation pattern consisting of an infinite number of infinitesimally thin lobes which we call (with a bit of imagination) `the hedge-hog' pattern; an extreme deformation of the isotropic radiation pattern with $G=1$ which we showed was optimal for $\eta>d$.
Interestingly, the hedge-hog pattern is by definition uniform in all orientations and therefore in some sense isotropic.

\section{Conclusions and Discussion \label{sec:conc}}

In this paper, we have investigated the connectivity properties of 3D ad hoc networks with randomly oriented anisotropically radiating nodes.
We have shown that for homogeneous systems (i.e. in the absence of boundary effects) the connectivity mass $\mathsf{M}$ is a key observable which characterises many important network properties: $i)$ the probability that two randomly selected nodes connect to form a pair $p_2$, $ii)$ the network mean degree $\mu$, $iii)$ the probability of an isolated node $d(0)$, and finally $iv)$ the probability of obtaining a fully connected network $P_{fc}$ at high node densities.
We therefore have focused on the explicit calculation of $\mathsf{M}$ for simple but practical antenna gain profiles (e.g. patch, dipole, and end-fire array antennas). 
Using the analytic expressions obtained, we have identified the ratio of spatial dimension $d$ to path loss $\eta$, as a key system parameter.
We have shown that when the antenna gain is concentrated on a small solid angle $\omega$, the connectivity mass $\mathsf{M}$ will scale as $\sim \omega^{2-2d/\eta}$.
Significantly, we have shown that for $\eta<d$, any directional deformation of the isotropic gain profile will increase $\mathsf{M}$ and therefore improve overall network connectivity.
In fact, we find that the more directional the gain, the better connected the network will be.
For $\eta>d$ however, all these observations are reversed and isotropic radiation leads to optimal network connectivity.
We have validated our results through Monte Carlo computer simulations of the network mean degree and have seen that border effects typically reduce the network mean degree - a feature particularly noticeable for highly directional radiation gains.

Random networks confined within a bounded domain are inhomogeneous systems and therefore boundary effects can have a significant impact on the network connectivity properties.
This is because nodes situated near the confinement boundary are likely to be of lower degree than those situated further away.
Therefore, the mean network degree is less than expected, particularly for highly directional gains which $a)$ may steer their main beam outside of the domain leading to so called blind spots, and $b)$ at low path loss exponents may have a characteristic connection range $r_0$ in their boresight direction which is greater than the typical domain size.
We have argued that these two effects have a greater impact for highly directional radiation patterns such as those of an end-fire array. 
Thus, in contrast to homogeneous systems, directionality in radiation gains is undesirable for networks operating in confined spaces, unless the network can be configured to eliminate the possibility of these eventualities.
To this end we have investigated multi-directional radiation patterns as a means to cover both bases (homogeneous and inhomogeneous systems).
We emphasize that the results presented in this paper are independent of the small-scale fading model used and therefore provide qualitative insight for wireless researchers and practitioners to consider in the future.


\section*{Acknowledgements}
\addcontentsline{toc}{section}{Acknowledgment}
The authors would like to thank the directors of the Toshiba Telecommunications Research Laboratory for their support.


\bibliographystyle{IEEEtran}
\bibliography{mybib}

\end{document}